\documentclass[twocolumn,showpacs,preprintnumbers]{revtex4}
\usepackage{graphicx}
\usepackage{bm}

\begin{document}

\title{Rationale for $10^{14}$ enhancement factor in single molecule Raman
spectroscopy}
\author{S.V. Gaponenko$^{1,}$}
\email{s.gaponenko@ifanbel.bas-net.by}
\author{D.V. Guzatov$^2$}
\affiliation{$^1$B.I.Stepanov Institute of Physics, NAS Belarus,
Minsk, 220072, Belarus \\ $^2$Scientific Research Center of Resource
Saving Problems, NAS Belarus, Grodno, 230023, Belarus}
\date{\today}

\begin{abstract}
We extend the Purcell's original idea [Phys. Rev. \textbf{69}, 682 (1946)] on
modification of photon spontaneous \textit{emission} rate to
modification of photon spontaneous \textit{scattering} rate. We find the interplay of
local incident field enhancement and local density of photon states
enhancement in close proximity to a silver nanoparticle may result in
up to $10^{14}$-fold rise of Raman scattering cross-section. Thus single
molecule Raman detection is found to be explained by consistent quantum
electrodynamic description without any chemical mechanism involved. A model of the so-called ``hot points'' in
surface enhanced spectroscopy has been elaborated as local areas with high
Q-factor at incident and scattered (emitted) light frequencies. For verification of the model we consider further
experiments including
transient Raman experiments to clarify incident field enhancement and
scanning near-field optical mapping of local density of photon states.
\end{abstract}

\maketitle

%\preprint{}

%\keywords{}

Since the discovery of molecular scattering of light with individual
signatures of specific bondings in 1928 \cite{ref1}, vibrational
spectroscopy has become the routine analytical tool in molecular physics and
chemistry. Discovery of the giant enhanced Raman
signals promoted by nanotextured metal surfaces and metal nanoparticles \cite%
{ref2} stimulated search for extreme Raman spectroscopy sensitivity and has
resulted in pioneering works \cite{ref3,ref4} reported on single molecule
Raman signatures. In spite of challenging experimental records, a consistent
theory explaining up to $10^{14}$ enhancement factors documented has not
been developed to date and the observation of single molecule Raman signals
remains unexplained. Typically, local incident field enhancement factor \cite%
{ref6} is considered as the major contribution to SERS signals giving
factors up to $10^{6}$ \cite{ref6} for most favorable combination of a metal
nanobody shape, a molecule location and incident light frequency. Further
enhancement factors are searched for among chemical mechanisms \cite{ref6}.
Notably, the theory is essentially reduced to classical electromagnetism
with no quantum electrodynamics (QED) involved.

A few years ago one the authors \cite{ref7} highlighted yet another
enhancement factor, namely local density of photon states effect on
Raman scattering rate in mesoscopic structures including metal
nanobodies.Indeed, Raman scattering rate $I$ (number of scattered
photons per second) can be written as the product of three terms

\begin{equation}
I\left( \omega ^{\prime }\right) =I_{0}\left( \omega \right) \left[ \text{%
interaction term}\right] D\left( \omega ^{\prime }\right) ,  \label{eq1}
\end{equation}

\noindent where $\omega $ is the incident light frequency, ${\omega }%
^{\prime }$ is the scattered light frequency, $I_{0}$ is incident
light intensity and $D({\omega }^{\prime })$ is the density of
photon states (photon DOS). In this presentation, the three
enhancement factors become apparent, i.e. local incident field
enhancement (the first term), chemical enhancement (the second term)
and density of photon states enhancement (the third term). Notably
this expression holds equally for Raman or Mandelstam-Brillouin
scattering ($\omega \neq \omega ^{\prime }$), for resonance
(Rayleigh) scattering ($\omega \equiv \omega ^{\prime }$) and for
spontaneous emission of photons. Such general formulation of
spontaneous emission and scattering of photons dates back to the
very first Dirac's paper on quantum electrodynamics \cite{ref8}.
While contribution of photon DOS redistribution is well recognized
and examined for spontaneous emission in mesoscopic structures (e.g.
\cite{ref9}-\cite{ref16} and refs therein), its contribution to
modified scattering of photons has not been systematically
recognized. The first calculation of local DOS contribution to SERS
for the case of a metal cylinder \cite{ref17} did offer an
optimistic value of $10^{7}$ which in fact should be taken as strong
overestimate since non-radiative contribution to decay rate has been
involved into consideration which does not contribute to photon
emission and scattering rate.

In this paper, we report on simultaneous consideration of incident field
enhancement and local density of photon states enhancement near a metal
particle with prolate spheroidal shape as a reasonable primary model for
single molecule Raman spectroscopy. Joint action of these two factors at the
same point of space is found to offer up to $10^{14}$-fold enhancement of
Raman scattering rate. To the best of our knowledge this is the first
evidence that consistent theory of single molecule Raman spectroscopy and
comprehensive description of so-called ``hot points'' in surface enhanced
spectroscopies can be constructed without necessarily involvement of
chemical mechanisms but with consistent QED consideration.

We start from the early expression for Raman scattering probability proposed
by G. Placzek \cite{ref18}

\begin{equation}
W_{RS} = \frac{\left( 2\pi \right) ^{2}\omega \omega ^{\prime }n}{\hbar ^{2}}%
\left\vert S\right\vert ^{2}\left[ n^{\prime }+\frac{\omega ^{\prime 2}}{%
\left( 2\pi c\right) ^{3}}\right] ,  \label{eq2}
\end{equation}

\noindent where $n$ is the incident photon number, ${n}^{\prime}$ is
the scattered photon number, $S$ is the matrix element of the
transition under consideration, $c$ is the speed of light in vacuum.
The first term in the square brackets describes stimulated
scattering whereas the second term corresponds to spontaneous
scattering. Notably, the second term is photon DOS in vacuum for a
given polarization within a unit solid angle. To get full
probability of scattering into $4\pi$ angle and 2 polarizations the
factor $8\pi$ should be added to arrive at full photon DOS in vacuum

\begin{equation}
D_{0}\left( \omega ^{\prime }\right) =\frac{\omega ^{\prime 2}}{\pi
^{2}c^{3}}.  \label{eq3}
\end{equation}

\noindent Therefore spontaneous Raman scattering rate in vacuum reads

\begin{equation}
W_{RS}=\frac{\pi \omega \omega ^{\prime }n}{2\hbar ^{2}}\left\vert
S\right\vert ^{2}D_{0}\left( \omega ^{\prime }\right) .  \label{eq4}
\end{equation}

\noindent For spontaneous scattering rate near a metal nanobody, the vacuum
density of photon states (\ref{eq3}) should be replaced by the \textit{local}
density of states (LDOS).

To calculate incident field enhancement of Raman scattering cross-section $%
\gamma _{RS}$ of a molecule near a nanoparticle normalized to that for the
same molecule in vacuum we use the well-known approach based on the
so-called electromagnetic theory of giant Raman scattering \cite{ref19}.
Within this theory we can consider a molecule with polarizability $\alpha $
located in z-axis of the Cartesian co-ordinates near a prolate
spheroid \cite{ref20}. If $\gamma _{RS}$ is considered as the function of
molecule position $z_{0}$ with other parameters fixed then the $\gamma
_{RS}\left( z_{0}\right) $ function will have maximum at certain point $%
z_{0}=b+\delta z$, where $b$ is a larger semi-axis of spheroid
(spheroid is stretched along z-axis). Generally, $\gamma _{RS}$ can
dramatically change with minor change in $\delta z$. For example,
shift in $\delta z$ by 1 {\AA } can result in 2-3 orders of the
magnitude change in $\gamma _{RS}$. For typical values of $\alpha $
= 10 $\text{\AA }^{3}$ for silver nanoparticles with 50-100 nm size
one has values of $\delta z\sim $ 1-2 {\AA }. Fig. 1 presents
$\gamma_{RS}$ as a function of relative frequency shift $\Delta \nu
=\left( \omega -\omega ^{\prime }\right) /\left( 2\pi \right) $ for
a molecule located near a silver prolate spheroidal nanoparticle
with $b$=80 nm and aspect ratio 8/5, and near silver nanosphere of
the same volume as spheroid at the point $z_{0}$. Dielectric
permittivity of silver from Ref. \cite{ref21} has been used in
calculations. The position of maximal values of Raman cross-section
is defined approximately by position of maximal absolute value of
nanoparticle's polarizability. For selected silver nanoparticles it
corresponds to 383.5 nm and 347.8 nm for spheroid, and 354.9 nm for
sphere. This defines selection of an incident light wavelength
chosen in the presented figure. For
normal orientation of induced dipole moment of a molecule the
enhancement factor $\gamma _{RS}$ readily reaches
$\sim 10^{11}$. In case of tangential orientation of induced dipole
moment such values are less than $\sim 10^{3}$. Normal orientation
of induced dipole moment implies $\mathbf{E}_{0}\parallel z$ whereas
tangential one means that $\mathbf{E}_{0}\parallel x$ or
$\mathbf{E}_{0}\parallel y$ axes, where $\mathbf{E}_{0}$ is an
electric field of an incident light.

\begin{figure}
\includegraphics[height=12 cm,angle=0]{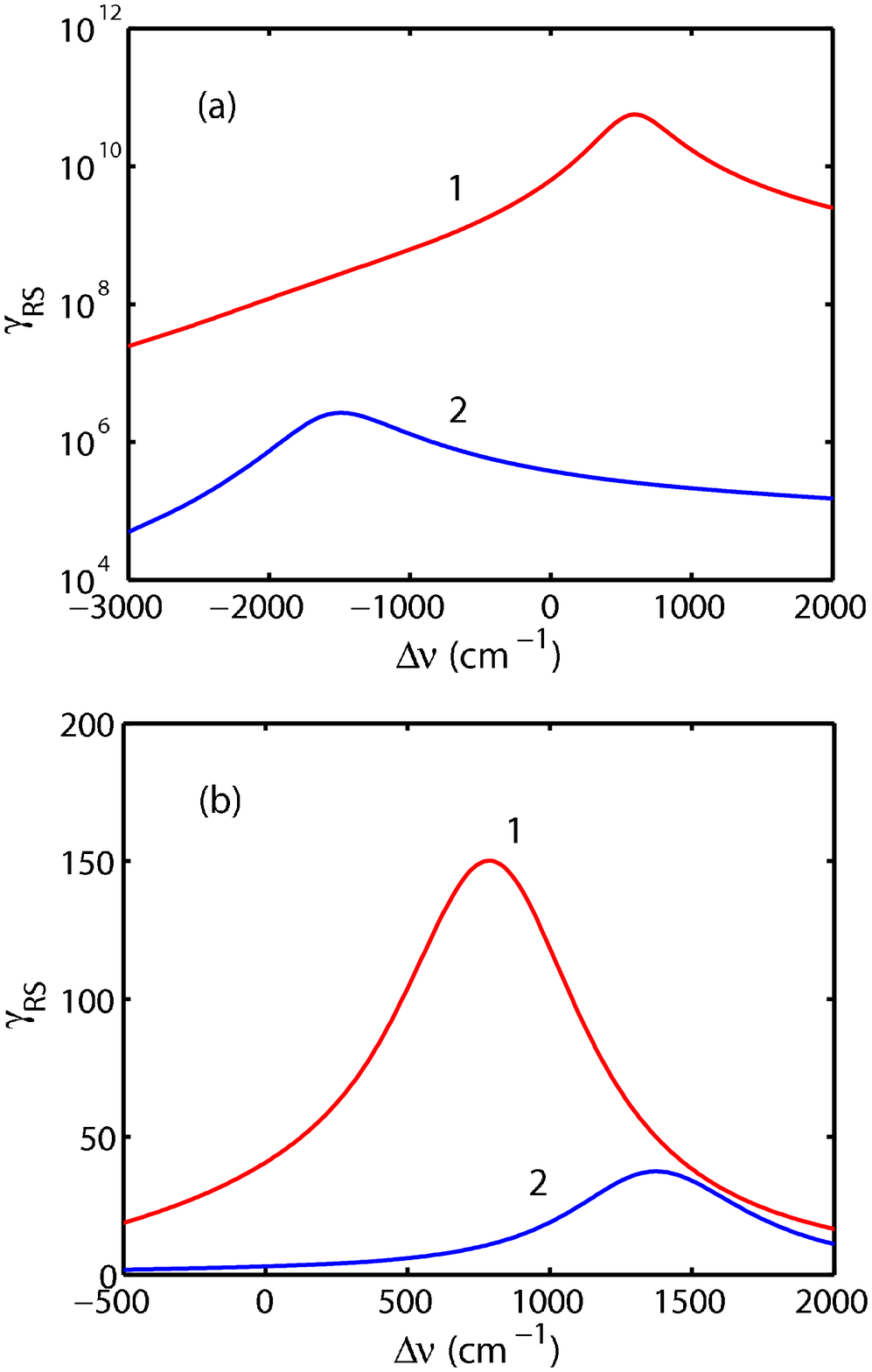}
\caption{(Color online) Relative Raman cross-section without photon LDOS
effect for a molecule with polarizability $\alpha$ = 10 $\text{\AA }%
^{3}$ located near a metal spheroidal nanoparticle (1) and
nanosphere (2) as function of spectral shift $\Delta \nu$. (a)
Incident field polarization is parallel to z-axis, wavelength of
incident light is 375 nm. (b) Incident field polarization is
parallel to x-axis, wavelength of incident light is 338 nm.}
\end{figure}

In spite of impressive factor from local incident field enhancement
it is well lower than observed $10^{14}$-fold single molecule Raman
signal \cite{ref3}. In what follows we show that further enhancement
can be understood if photon LDOS near a metal
nanobody is properly taken into account as is seen from Eq.
(\ref{eq4}). Photon LDOS near a nanobody $D$ normalized with respect
to that in free space $D_{0}$ formally coincides with the
enhancement factor radiative rate gains near a nanobody with respect
to rate in free space. This statement can be proved based on the
fluctuation-dissipation theorem \cite{ref22} and is proposed as
reasonable operational approach to the very definition of local
density of photon states \cite{ref23}. With modified LDOS Raman
scattering cross-section reads

\begin{equation}
\gamma _{RS}^{LDOS}\left( \omega ,\omega ^{\prime }\right) =\gamma
_{RS}\left( \omega ,\omega ^{\prime }\right) \frac{D\left( \omega ^{\prime
}\right) }{D_{0}\left( \omega ^{\prime }\right) }.  \label{eq9}
\end{equation}

Calculation of LDOS has been made based on solution of a
quasi-steady-state problem for a dipole electromagnetic field source
near a nanoellipsoid in accordance with our previous works
\cite{ref28}. In a quasi-steady-state approximation the full rate of
spontaneous decay can be partitioned into two components. The first
one is the radiative part related to photon emission. The second one
is the non-radiative part related to dissipation of energy to a
metal nanobody. The radiative part is found as the ratio of emmitted
power by a dipole source near a nanobody to power of the same source
in free space. Unlike \cite{ref17} only radiative part of the full
decay rate has been accounted for since it is the radiative part of
the full atomic decay rate which is determined by LDOS. The
non-radiative part corresponds to Joule losses (heating of a
nanobody) \cite{ref24}. In Fig. 2 calculated Raman cross-section is
plotted with Eq. (\ref{eq9}) taken into account as function of
spectral shift for the same molecule and nanoparticle parameters and
their relative displacement as in Fig. 1. One can see in addition to
local incident field enhancement, local density of states
enhancement provide factors of the several orders of the magnitude.
Notably, maximal enhancement occurs for detuning from the incident
field enhancement which is typical Raman shift for common organic
molecules. Optimal combination of the two enhancement factors can
result in more than $10^{14}$-fold enhancement factors for proper
displacement of a molecule near a nanoparticle and for proper
orientation of its dipole moment with respect to incident field and
a spheroid axis (Fig. 2a). Such values inherent in certain ``hot
points'' make Raman detection of single molecules plausible.
Enhancement is essentially frequency dependent which qualitatively
corresponds to experiments but often is attributed to chemical
factors.

\begin{figure}
\includegraphics[height=12 cm,angle=0]{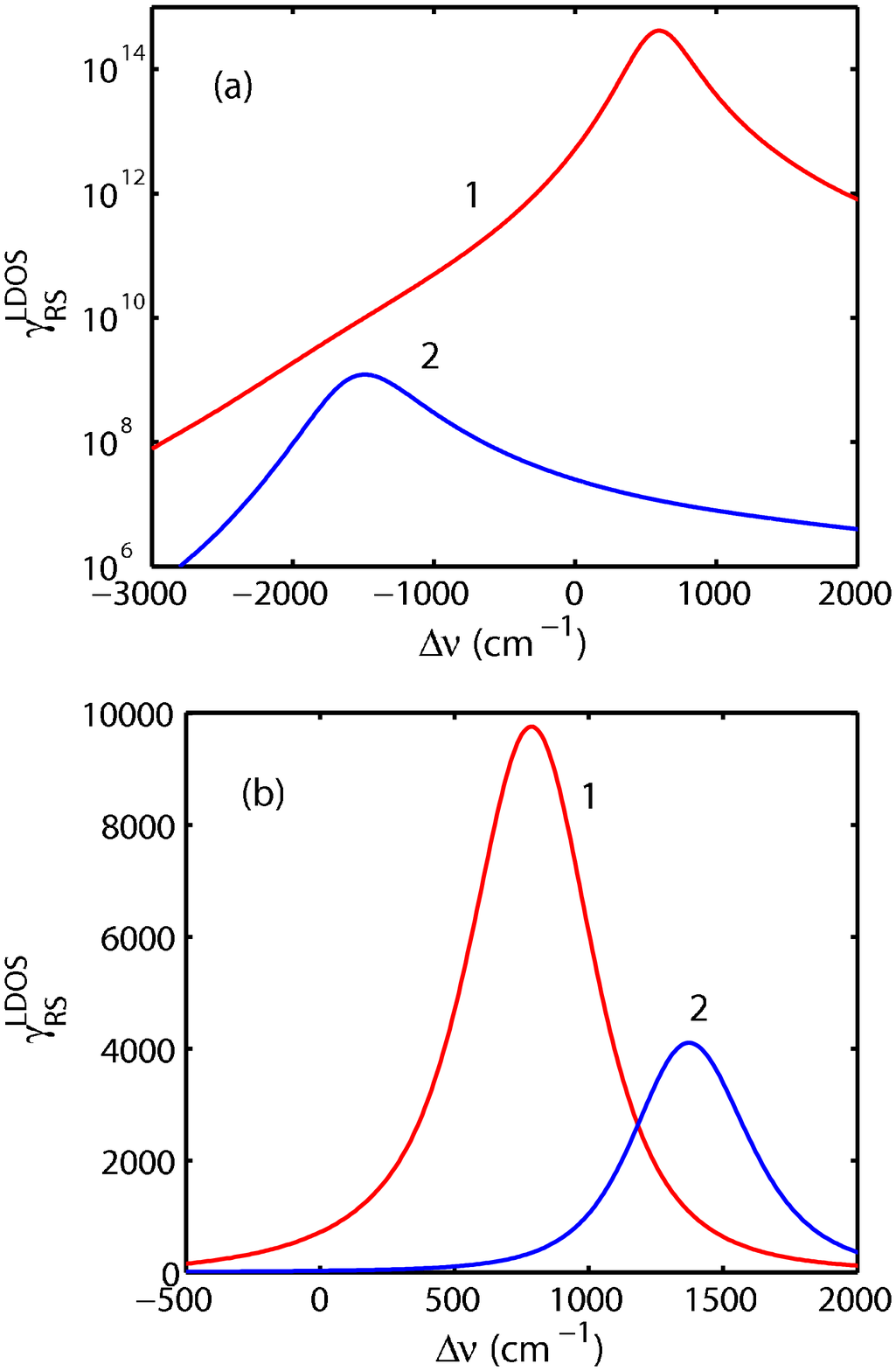}
\caption{(Color online) Relative Raman cross-section with photon
LDOS effect for a molecule with polarizability $\alpha$ = 10
$\text{\AA }^{3}$ located near a metal spheroidal nanoparticle (1)
and nanosphere (2) as function of spectral shift $\Delta \nu$. (a)
Incident field polarization is parallel to z-axis, wavelength of
incident light is 375 nm. (b) Incident field polarization is
parallel to x-axis, wavelength of incident light is 338 nm.}
\end{figure}

The above consideration is valid not only for molecular spectroscopy
but for all versions of vibrational spectroscopies, e.g. it can be
applied for single quantum dot vibrational spectroscopy
\cite{ref25}. It is also valid for Mandelstam-Brillouin scattering
as well as for Rayleigh scattering. The latter has been discussed in
our previous paper \cite{ref26} and gains additional argumentation
in the context of the recent report on enhanced hyper-Rayleigh
scattering in metal-dielectric nanostructures \cite{ref27}.

We believe that simultaneous action of incident field enhancement and local
density of photon states enhancement does provide a reasonable rationale for
single molecule Raman spectroscopy. The results presented in Fig. 2 are
considered as a first step towards extensive theory for single molecule
Raman detection. Triaxial ellipsoidal nanoparticles are expected to offer
even higher field enhancement and LDOS enhancement factors \cite{ref28}.
Furthermore, coupled metal nanoparticles which have been found to exhibit
higher efficiency in Raman scattering enhancement \cite{ref29} have also
been proven theoretically to possess superior local field \cite{ref29} and
LDOS \cite{ref30} enhancement in the spacing between spheres and are
believed their SERS efficiency can be described by simultaneous incident
field and LDOS enhancements.

The proposed model sheds light on the so-called \textquotedblleft hot
points\textquotedblright\ as such places on a nanotextured metal surface or
near metal nanobodies where simultaneous spatial redistribution of
electromagnetic field occurs both at the frequency of the incident radiation
$\omega $ and at the frequency of scattered radiation $\omega ^{\prime }$.
The first effect is the so-called field enhancement factor whereas the
second is local density of states enhancement. Enhancement of photon LDOS
starting from the pioneering paper by E.M. Purcell \cite{ref9} can be
interpreted as development of the certain Q-factor in the space region where
a test emitter (atom or other quantum system) is placed. Since Q-factor
implies a system is capable to accumulate energy (then Q value equals to the
ratio of energy accumulated in the system to the portion of energy the
system looses in a single oscillation period), formation of high local
density of states areas in many instances can be treated as development of
multiple microcavities at the scattered frequency over nanotextured metal
surface. From the other side, such microcavities promote electromagnetic
wave tunneling including light leakage towards the surface in near-field
optical microscopy. Therefore surface mapping of high LDOS
areas can be performed by scanning near-field microscopy as has
been proposed in Ref. \cite{ref31} but to the best of the authors' knowledge
has never been applied to SERS-active structures.

Local field enhancement for incident light can not be interpreted as
surface redistribution of incident light, i.e. as
light ``microfocusing'' as anticipated by many authors. Since SERS
is considered within linear light-matter
interaction (contrary to e.g. surface enhanced second harmonic
generation) the total Raman signal harvesting from a piece of area
containing statistically large number of molecules will be the same
independently of surface redistribution of light intensity because
total incident light intensity integrated over the piece of area
remains the same. Within the framework of linear light-matter
interaction, Raman signal enhancement by means of incident field
enhancement can only be understood in terms of high local Q-factors
for incident light, i.e. in terms of light \textit{accumulation near
the surface} rather than light \textit{redistribution over the
surface}. Q-fold rise up of light intensity then occurs near hot
points as it happens in microcavities and interferometers. However,
accumulation of light energy needs certain time. Therefore huge
Raman signals can develop only after certain time which is necessary
for transient processes to finish resulting in steady increase of
incident light intensity near hot point. Transient SERS experiments
are therefore to be performed to clarify Q-factor effects in hot
points formation.

Local DOS enhancement in a sense accounts for concentration of
electromagnetic field at $\omega ^{\prime }$. This statement
unambiguously implies \textit{probe}, non-existing field
\cite{ref32}. However, concentration of real field by many authors
was anticipated to offer $\left\vert \mathbf{E}\left( \omega
^{\prime }\right) \right\vert ^{2}$ enhancement factor \cite{ref6}
by analogy to $\left\vert \mathbf{E}_{0}\left( \omega \right)
\right\vert ^{2}$ factor for input light intensity. That
anticipation is by no means justified because $\left\vert
\mathbf{E}\left( \omega ^{\prime }\right) \right\vert ^{2}$
enhancement occurs only in the close subwavelength-scale vicinity of
a nanobody and can not contribute to light harvesting in typical far
field experiments. Concentration of really emitted light
can actually contribute to SERS but only
as \textit{induced} Raman scattering [$n^{\prime }$ term in Eq.
(\ref{eq2})].

In conclusion, a rationale has been proposed for more than $10^{14}$-fold
enhancement factors in Raman spectroscopy in terms of local field
enhancement and local density of photon states enhancement in the same point
but at different frequencies, a model of the so-called ``hot points'' has
been elaborated as local areas with high Q-factor at incident and scattered
light frequencies and further experiments towards verification of the model
have been outlined. The proposed consideration extends the original
Purcell's idea on strong modification of photon spontaneous \textit{emission}
rate to modification of spontaneous photon \textit{scattering} rates.

The work has been supported by the EU NoE ``PHOREMOST'', National
Research Program ``Crystalline and Molecular Structures'' and by
Belarusian National Basic Research Foundation.

\end{document}